\documentclass[pra,twocolumn,amsmath,amssymb,showpacs,longbibliography]{revtex4-1}
\usepackage{hyperref}
\pdfoutput=1
\usepackage{graphicx}

\usepackage{amsmath}
\usepackage{bm}
\usepackage{amssymb}
\usepackage{textcomp}
\usepackage{stmaryrd} 
\usepackage{ulem}

\usepackage{import}
\usepackage{color}
\usepackage{verbatim}

\usepackage{float}

\makeatletter
\DeclareRobustCommand{\element}[1]{\@element#1\@nil}
\def\@element#1#2\@nil{%
  #1%
  \if\relax#2\relax\else\MakeLowercase{#2}\fi}
\makeatother

\newcommand{\tr}[1]{{\textrm{#1}}}

\newcommand\abs[1]{\left| #1 \right|}

\def\XXint#1#2#3{{\setbox0=\hbox{$#1{#2#3}{\int}$}
     \vcenter{\hbox{$#2#3$}}\kern-.5\wd0}}

\begin{document}

\title{Direct versus indirect band gap emission and exciton-exciton annihilation in atomically thin molybdenum ditelluride (MoTe$_2$)}

\author{Guillaume Froehlicher}
\affiliation{Institut de Physique et Chimie des Mat\'eriaux de Strasbourg and NIE, UMR 7504, Universit\'e de Strasbourg and CNRS, 23 rue du L\oe{}ss, BP43, 67034 Strasbourg Cedex 2, France}

\author{Etienne Lorchat}
\affiliation{Institut de Physique et Chimie des Mat\'eriaux de Strasbourg and NIE, UMR 7504, Universit\'e de Strasbourg and CNRS, 23 rue du L\oe{}ss, BP43, 67034 Strasbourg Cedex 2, France}

\author{St\'ephane Berciaud}
\email{stephane.berciaud@ipcms.unistra.fr}
\affiliation{Institut de Physique et Chimie des Mat\'eriaux de Strasbourg and NIE, UMR 7504, Universit\'e de Strasbourg and CNRS, 23 rue du L\oe{}ss, BP43, 67034 Strasbourg Cedex 2, France}

\begin{abstract}
We probe the room temperature photoluminescence of $N$-layer molybdenum ditelluride (MoTe$_2$) in the continuous wave (cw) regime. The photoluminescence quantum yield of monolayer MoTe$_2$ is three times larger than in bilayer MoTe$_2$ and forty times greater than in the bulk limit. Mono- and bilayer MoTe$_2$ display almost symmetric emission lines at $1.10~\rm eV$ and $1.07~\rm eV$, respectively, which predominantly arise from direct radiative recombination of the A exciton. In contrast, $N\geq3\tr{-layer}$ MoTe$_2$ exhibits a much reduced photoluminescence quantum yield and a broader, redshifted and seemingly bimodal photoluminescence spectrum. The low- and high-energy contributions are attributed to emission from the indirect and direct optical band gaps, respectively. Bulk MoTe$_2$ displays a broad emission line with a dominant contribution at 0.94~eV that is assigned to emission from the indirect optical band gap. As compared to related systems (such as MoS$_2$, MoSe$_2$, WS$_2$ and WSe$_2$), the smaller energy difference between the monolayer direct optical band gap and the bulk indirect optical band gap leads to a  smoother increase of the photoluminescence quantum yield as $N$ decreases.  In addition, we study the evolution of the photoluminescence intensity in monolayer MoTe$_2$ as a function of the exciton formation rate $W_\tr{abs}$ up to $3.6\times 10^{22}~\rm cm^{-2} s^{-1}$. The lineshape of the photoluminescence spectrum remains largely independent of $W_\tr{abs}$, whereas the photoluminescence intensity grows sub-linearly above $W_\tr{abs}\sim 10^{21}~\rm cm^{-2} s^{-1}$. This behavior is assigned to exciton-exciton annihilation and is well-captured by an elementary rate equation model. 

\end{abstract}


\maketitle

\section{Introduction}

Transition metal dichalcogenides~\cite{Wilson1969} (herein denoted MX$_2$, where M=Mo, W, Re and X=S, Se, Te) are an actively investigated class of layered materials, whose basic electronic, optical and vibrational properties depend critically on the number of layers $N$ that compose a given sample~\cite{Mak2010,Splendiani2010,Lee2010,Molina2011,Froehlicher2015}. $N$-dependent properties are remarkably illustrated by the transition from indirect optical band gap, in the bulk form, to direct optical band gap \footnote{Throughout the text, ``direct optical band gap'' and ``indirect optical band gap'' will thus denote the energy of the photons emitted from the direct and indirect band-edge excitons, respectively.} at monolayer thickness  that occurs in $2Hc$ Mo- and W-based semiconducting MX$_2$~\cite{Mak2010,Splendiani2010,Tongay2012,Tonndorf2013,Zhao2012,Cappelluti2013}. Direct optical band gaps, together with the possibility of achieving valley polarization for resonantly pumped band-edge excitons in monolayer MX$_2$~\cite{Xu2014}, open original perspectives for two-dimensional optoelectronics~\cite{Wang2012} and valleytronics~\cite{Mak2014}. 

An interesting direction in this field, consists in exploring MX$_2$ with smaller optical band gaps (\textit{i.e.}, related to the formation of tightly bound excitons~\cite{Molina2013,Qiu2013,Ramasubramaniam2012,Chernikov2014,He2014,Ye2014,Wang2015,Zhu2015,Li2015}) than the extensively studied monolayers of MoS$_2$, MoSe$_2$, WS$_2$, WSe$_2$, whose optical band gaps lie in the range $1.5-2.0$~eV~\cite{Li2015}. Such endeavors are motivated by the possibility of achieving gate-controlled ambipolar transport more easily~\cite{Lezama2014,Lin2014} and to extend optoelectronic applications of MX$_2$ and related van der Waals heterostructures~\cite{Geim2013} into the near-infrared range.  Among possible candidates, $N$-layer molybdenum ditelluride (MoTe$_2$)~\cite{Lezama2014,Lin2014,Ruppert2014,Lezama2015,Yang2015,Yamamoto2014,Froehlicher2015,Chen2015b,Koirala2016,Kuiri2016}, as well as rhenium diselenide (ReSe$_2$) have emerged very recently. While $N$-layer ReSe$_2$ crystals exhibit a distorted $1T$-phase~\cite{Tongay2014,Zhao2015,Wolverson2014,Lorchat2016} and are indirect optical band gap semiconductors, irrespective of $N$~\cite{Zhao2015}, stable $N$-layer $2Hc$-MoTe$_2$ crystals have been shown to undergo a transition from indirect (for bulk MoTe$_2$) to direct (for monolayer MoTe$_2$) optical band gap~\cite{Ruppert2014,Lezama2015}. However, the exact value of $N$ at which the crossover occurs is a matter of debate~\cite{Lezama2015} and a detailed analysis of the photoluminescence (PL) lineshape in $N$-layer MoTe$_2$ is still lacking. In addition, the evolution of the PL spectrum and integrated PL intensity of monolayer  MoTe$_2$  with increasing exciton density remains unexplored so far.

In this article, we address the room temperature PL properties of $N$-layer $2Hc$-MoTe$_2$ in the continuous wave (cw) regime. Our data show that the PL quantum yield of monolayer MoTe$_2$ is approximately three times (forty times) larger than that of bilayer (bulk) MoTe$_2$, confirming the transition from a bulk indirect optical band gap (giving rise to an emission line at $0.94~\tr{eV}$) to a direct optical band gap at $1.10~\tr{eV}$~\cite{Ruppert2014}. Moreover, an analysis of the PL lineshapes reveals two close-lying contributions to the PL spectra. For mono- and bilayer MoTe$_2$, the observation of similar, almost symmetric PL spectra indicates that the crossover from dominant indirect to dominant direct band gap emission presumably occurs between $N=3$ and $N=2$ at room temperature.  For $N=3$ to $N=7$ layers MoTe$_2$, the low- and high-energy PL features are assigned to emission from the indirect and direct optical band gaps, respectively. Finally, the  PL intensity of monolayer MoTe$_2$ levels off with increasing laser intensity (\textit{i.e.,} as the exciton formation rate increases). This non-linear behavior unveils the critical role of exciton-exciton annihilation in atomically thin MoTe$_2$, as also reported recently in other MX$_2$~\cite{Sun2014,Kumar2014,Mouri2014,Zhu2015,Yuan2015,Yu2016}.

\section{Methods}

$N$-layer crystals of trigonal prismatic ($2Hc$ phase) MoTe$_2$ (hereafter denoted MoTe$_2$, see Fig.~\ref{Fig1}(a)) were prepared by mechanical exfoliation of commercially available bulk crystals (2D semiconductors) onto Si wafers covered with a 90~nm-thick SiO$_2$ epilayer (see Fig.~\ref{Fig1}(b)).  The number of layers was first estimated from optical contrast and further confirmed by ultralow-frequency micro-Raman spectroscopy (see Fig.~\ref{Fig1}(c)-(d)). PL and Raman spectra were recorded in ambient conditions, both in a backscattering geometry, using a home-built setup. In Raman experiments, a combination of one narrow bandpass filter and two narrow notch filters (Optigrate) was used in order to attain the low-frequency range of the spectrum. After optimization, Raman features at frequencies as low as $4.5~\tr{cm}^{-1}$ could be measured (see Fig.~\ref{Fig1}(c)).
In all experiments, freshly prepared samples \cite{Chen2015b} were optically excited using a single longitudinal mode, linearly polarized, 2.33~eV (532~nm) laser beam focused onto a $\approx\,600~\tr{nm}$-diameter spot using a high numerical aperture objective (NA=0.65). PL spectra in Figs.~\ref{Fig2}-\ref{Fig3} were recorded in the linear regime at a laser intensity of approximately~$1.5~\rm kW/cm^2$, using a single monochromator equipped with a 150 grooves/mm ruled grating coupled to a thermoelectrically cooled two-dimensional InGaAs array (Princeton Instruments NIRvana). Raman spectra were recorded at a laser intensity of approximately 60~$\rm kW/cm^2$, using the same monochromator equipped with a 2400 grooves/mm holographic grating, coupled to a two-dimensional liquid nitrogen cooled charge-coupled device (CCD) array. We have verified that the higher laser intensities employed for Raman studies were not damaging our samples. 



\section{Determination of the number of layers $N$}

\begin{figure}[!tbh]
\begin{center}
\includegraphics[width=1\linewidth]{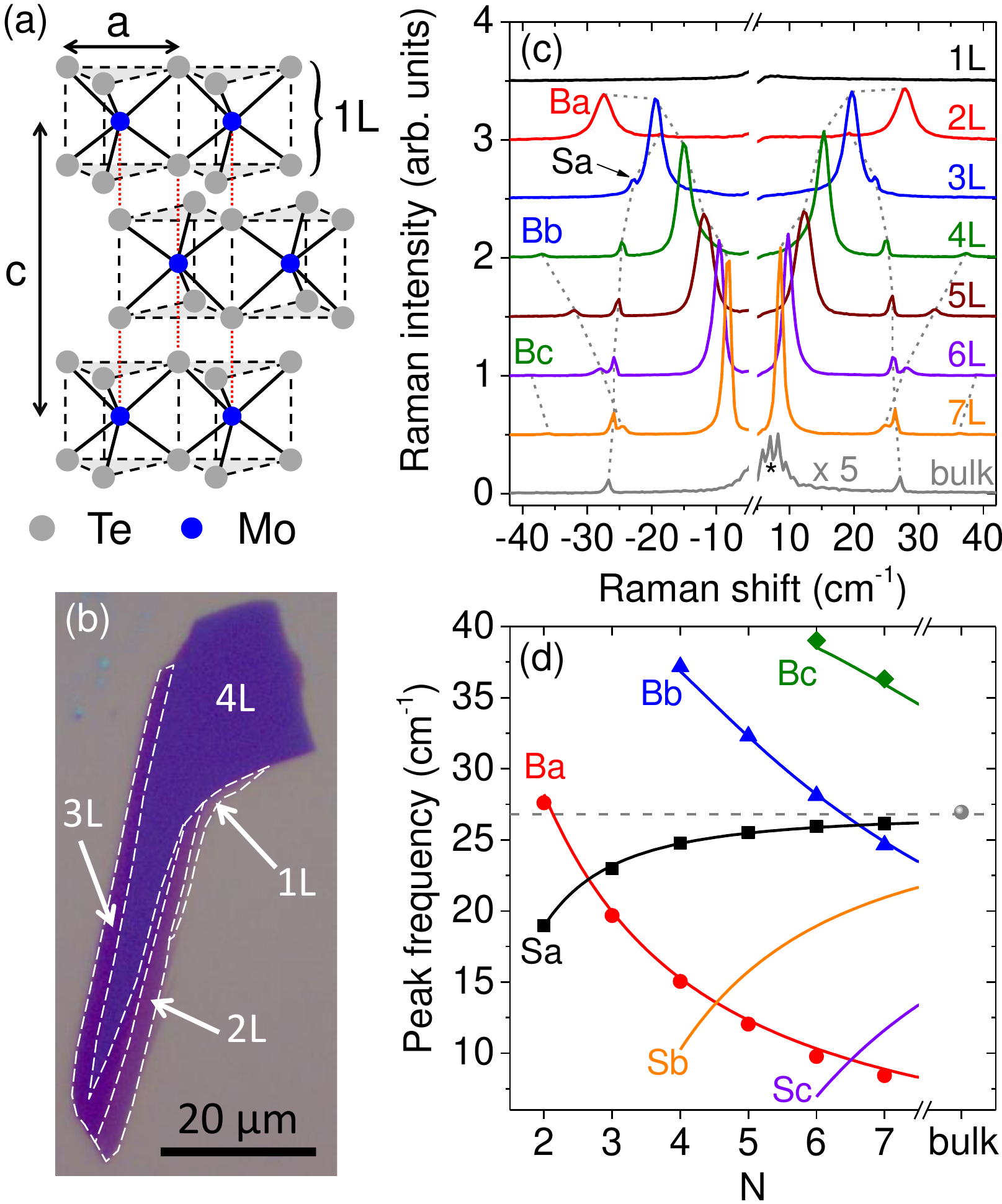}
\caption{(a) Side view of the crystal structure of $2Hc$-MoTe$_2$. (b) Optical image of a MoTe$_2$ flake (deposited onto a Si/SiO$_2$ substrate) containing mono to tetralayer domains. The boundaries of the various layers are highlighted with dashed lines. (c) Ultralow-frequency Raman spectra of $N=1$ to $N=7$ layer MoTe$_2$ and of bulk MoTe$_2$. The asterisk highlights residual contributions from the exciting laser beam. (d) Fan diagram of the interlayer shear (Sa, Sb and Sc) and breathing (Ba, Bb and Bc) modes of MoTe$_2$. Symbols are frequencies extracted from the Raman spectra in (c). The solid lines are theoretical calculations based on a linear chain model and the gray dashed line corresponds to the bulk frequency of the interlayer shear mode.}
\label{Fig1}
\end{center}
\end{figure}

Figure~\ref{Fig1}(c) shows the low-frequency Raman spectra (in the range $0 - 40~\tr{cm}^{-1}$) of $N$-layer MoTe$_2$, from $N=1$ to $N=7$, and of a thick sample ($N\gtrsim50$ layers) considered as a bulk reference. As previously reported~\cite{Froehlicher2015}, the low-energy features observed for $N\geq2$ correspond to \textit{interlayer} shear (LSM) and breathing (LBM) modes (see the gray dashed lines in Fig.~\ref{Fig1}(c)).  In bulk MoTe$_2$, the LBM is silent~\cite{Michel2012,Zhao2013} and only a single peak, assigned to the LSM can be observed.  The evolution of the LSM and LBM frequencies with $N$ can be analytically described by the expression $ \omega_k(N)=\frac{\omega_0}{\sqrt{2}}\sqrt{1-\cos\left(\frac{k\pi}{N}\right)}$ (with $k=1,...,N-1$) deduced from a finite linear chain model~\cite{Tan2012,Zhao2013,Zhang2013,Boukhicha2013,Froehlicher2015}. Using this expression, the observed modes were fit using $k=N-1$ for the LSM branch (Sa) and $k=1,3,5$ for the LBM branches (Ba, Bb and Bc, respectively), as shown in Fig.~\ref{Fig1}(d). These fits yield bulk frequencies $\omega^\tr{LSM}_0=26.8~\tr{cm}^{-1}$ and $\omega^\tr{LBM}_0=39.9~\tr{cm}^{-1}$ in excellent agreement with the results in Ref.~\cite{Froehlicher2015}, further confirmed in Refs.~\cite{Grzeszczyk2016,Song2016}. This analysis permits an unambiguous determination of $N$.

\section{PL Spectra of $N$-layer \element{Mo}\element{Te}$_2$}

\begin{figure*}[!th]
\begin{center}
\includegraphics[width=1\linewidth]{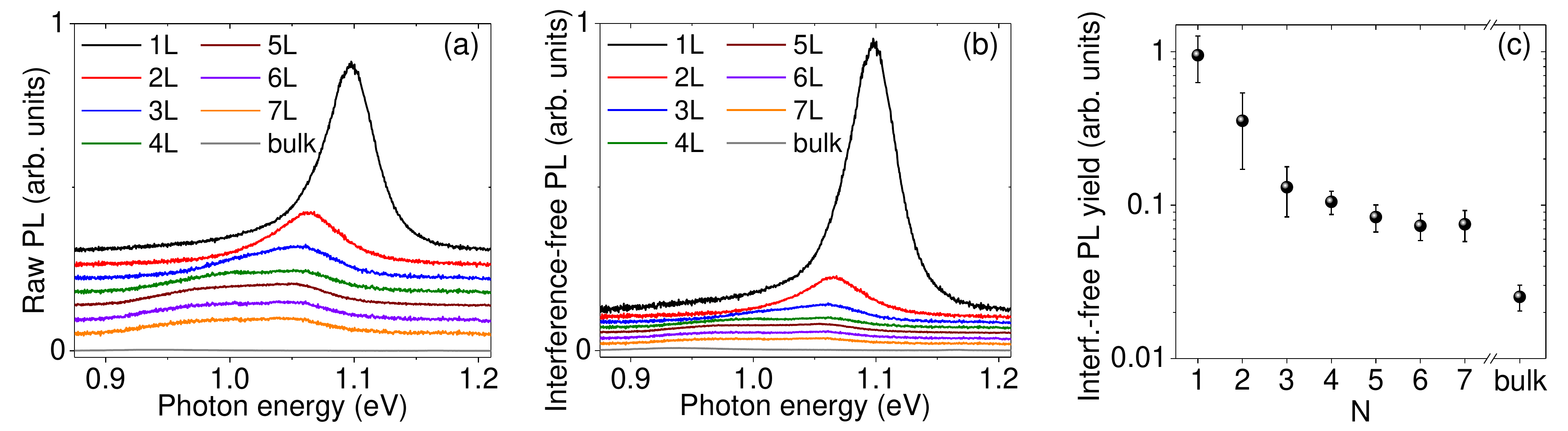}
\caption{(a) Raw and (b) \textit{interference-free} photoluminescence spectra of $N=1$ to $N=7$ layer MoTe$_2$ and of bulk MoTe$_2$ deposited on a Si/SiO$_2$ substrate. (c) Average total integrated intensities of the \textit{interference-free} photoluminescence spectra as a function of $N$ obtained on three samples (except for $N=5$ and $N=6$, for which only one sample was studied).}
\label{Fig2}
\end{center}
\end{figure*}

Figure~\ref{Fig2}(a) displays the raw PL spectra of the MoTe$_2$ samples ($N=1$ to $N=7$ and bulk) previously introduced in Fig.~\ref{Fig1}. It is well known that interference effects strongly affect the exciton formation rate, as well as the Raman~\cite{Yoon2009,Li2012b} and PL~\cite{Buscema2014} response of layered materials deposited on layered substrates such as Si/SiO$_2$. In order to take these phenomena into account, \textit{interference-free} PL spectra were obtained by normalizing the raw spectra by the enhancement factor calculated following Refs.~\cite{Yoon2009,Li2012b} (see Fig.~\ref{Fig2}(b) and Supplemental Material~\cite{SMnote}). This procedure allows us to compare, in Fig.~\ref{Fig2}(c), the \textit{interference-free} PL quantum yields, which are proportional to the integrated intensity of the \textit{interference-free} PL spectra. Note that the enhancement factor takes into account the number of layers and is thus homogeneous to a length. Therefore, the \textit{interference-free} PL quantum yields are given per unit length. Moreover contrary to what was reported in Ref.~\cite{Yang2015}, the PL background from the Si substrate is negligible in our experiments (see Supplemental Material~\cite{SMnote}).

Figure~\ref{Fig2}(b) displays the \textit{interference-free} PL spectra. The PL lineshapes are marginally affected as compared to the raw spectra, whereas the integrated \textit{interference-free} PL intensities are significantly modified. As $N$ increases, we immediately notice that (i) the integrated PL intensity decreases monotonically and is three (resp. forty) times smaller in bilayer (resp. bulk) MoTe$_2$ than in the monolayer limit (see Fig.~\ref{Fig2}(c)), (ii) the PL peak energy redshifts from 1.10~eV at monolayer thickness down to 0.94~eV in the bulk limit and (iii) the PL lineshapes are slightly asymmetric for $N=1,2$ and clearly bimodal for $N\geq3$.  The first two observations are consistent with a transition from an indirect optical band gap in the bulk limit to a direct optical band gap for $N=1$~\cite{Ruppert2014}. The increase in PL quantum yield as $N$ decreases is moderate, as compared to recent observations in MoS$_2$, MoSe$_2$, WS$_2$, and WSe$_2$~\cite{Mak2010,Tonndorf2013,Zhao2012}. This behavior is due to the smaller energy difference between the bulk emission from the indirect optical band gap and the direct optical band gap. For instance, the latter is approximately 0.6~eV in MoS$_2$~\cite{Mak2010} and 0.5~eV in MoSe$_2$~\cite{Tonndorf2013}.

\section{Indirect-to-direct optical band gap crossover}

\begin{figure}[!htb]
\begin{center}
\includegraphics[width=0.95\linewidth]{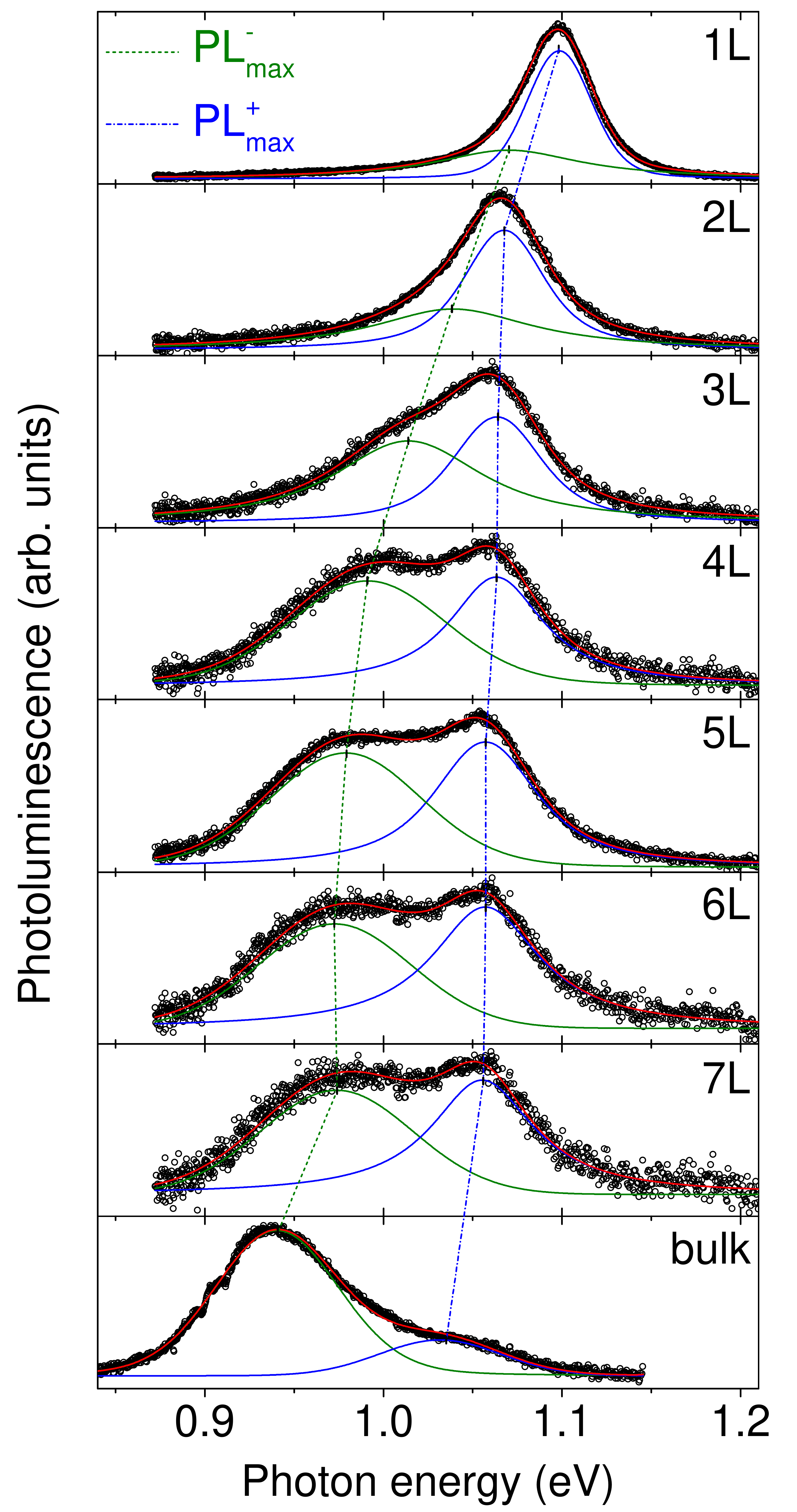}
\caption{Normalized \textit{interference-free} photoluminescence spectra of $N=1$ to $N=7$ layer MoTe$_2$ and of bulk $\rm{MoTe}_2$. The spectra are the same as in Fig.~\ref{Fig2}(b). The data (black open circles) are fit using the sum of two Voigt profiles (red solid lines). The green and blue solid lines are the PL$^{-}$ and PL$^{+}$ features, respectively. The green dotted and blue dash-dotted lines mark the evolution of the associated peak energies, denoted $\rm PL_{max}^{-}$ and $\rm PL_{max}^{+}$, respectively, as a function of the number of layers.}
\label{Fig3}
\end{center}
\end{figure}

\begin{figure}[!tb]
\begin{center}
\includegraphics[width=1\linewidth]{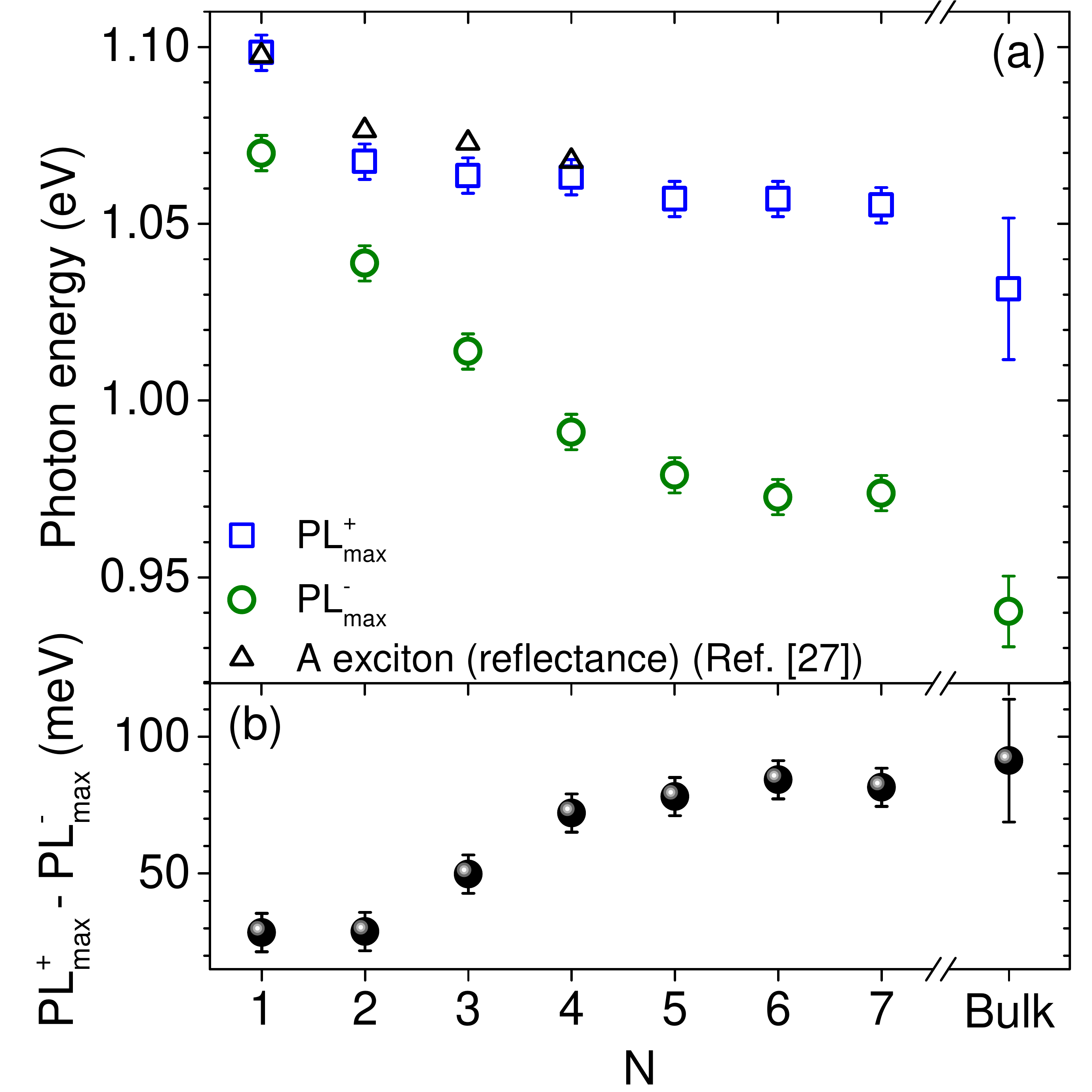}
\caption{(a) Energies of the photoluminescence peaks PL$^-_\tr{max}$ (green open circles) and PL$^+_\tr{max}$ (blue open circles)  as a function of the number of layers $N$. The data are extracted from the fits shown in Fig.~\ref{Fig3} and correspond to the same samples as in Fig.~\ref{Fig2}(c). Our experimental measurements are compared to the reflectance  measurements from Ref.~\cite{Ruppert2014} (open black triangles). (b) Energy difference between the two photoluminescence peaks as a function of the number of layers $N$. }
\label{Fig4}
\end{center}
\end{figure}

The exact value of $N$ at which the crossover occurs is still debated. At room temperature, Ruppert \textit{et al.}~\cite{Ruppert2014} have suggested a crossover when reaching the monolayer limit, while at low temperature ($4 - 180~\tr{K}$),  Lezama \textit{et al.}~\cite{Lezama2015} concluded that the crossover occurs between $N=3$ and $N=2$. Very recently, at 10~K, Robert \textit{et al.}~\cite{Robert2016b} have observed similar PL intensities in mono- and bilayer MoTe$_2$ and a slightly longer PL decay time in  bilayer MoTe$_2$ than in monolayer MoTe$_2$, suggesting that PL in bilayer MoTe$_2$ may in part originate from the direct optical band gap. However, there is no apparent contradiction between these claims since it is well-known that temperature might affect the crossover~\cite{Tongay2012}. 
Here, we could clearly identify two subfeatures within each PL spectrum, as illustrated in Fig.~\ref{Fig3}. We may now wonder whether these two contributions may be associated with the direct and indirect optical band gaps. To answer this question, we have systematically fit the PL spectra with a double Voigt profile (see Fig.~\ref{Fig3}) and extracted the high- (PL$^+$) and low-energy (PL$^-$) contributions. Figure~\ref{Fig4} displays the peak positions PL$^+_\tr{max}$ and PL$^-_\tr{max}$. 

First, the PL spectrum of monolayer MoTe$_2$ exhibits an almost symmetric lineshape dominated by a relatively narrow PL$^+$ feature with a full width at half maximum (FWHM) of approximately 50~meV. The peak position PL$^+_\tr{max}$ matches the energy of the A exciton measured by room temperature differential reflectance spectroscopy by Ruppert \textit{et al.}~\cite{Ruppert2014} (see Fig.~\ref{Fig4}(a)) and PL$^+_\tr{max}$ is therefore identified as the direct optical band gap energy. The PL$^-$ shoulder is much broader (FWHM of approximately 100~meV) and has lower integrated intensity than that of the PL$^+$ peak. Assuming that monolayer MoTe$_2$ is a direct optical band gap semiconductor, the PL$^-$ feature cannot arise from the indirect optical band gap. Since the energy difference between the $\tr{PL}^{\pm}$ features is approximately $30~\tr{meV}$ (see Fig.~\ref{Fig4}(b)), the PL$^-$ peak can tentatively be assigned to emission from charged A excitons (\textit{i.e.}, trions~\cite{Lezama2015,Yang2015}) or to exciton-phonon sidebands involving coupling of A excitons with $\Gamma$-point optical phonons (whose energies lie in the range 15-35~meV~\cite{Yamamoto2014,Froehlicher2015}). 

Second, the PL spectrum of bilayer MoTe$_2$ is slightly redshifted (by about 30~meV) with respect to the monolayer case, with a normalized PL quantum yield about three times smaller than that of monolayer MoTe$_2$, suggesting that bilayer MoTe$_2$ is not a direct optical band gap semiconductor. However, although the bilayer PL spectrum  is appreciably broader than that of the monolayer PL spectrum (FWHM of approximately 65~meV), the spectra are similar. Indeed, PL$^+_\tr{max}$ also matches the energy of the A exciton for $N=2$~\cite{Ruppert2014}. In addition, the PL$^+$ peak is more intense than the PL$^-$ peak, and the energy difference between the peak positions of these two features remains approximately $30~\tr{meV}$ (see Fig.~\ref{Fig4}), as in monolayer MoTe$_2$. These observations indicate that the room temperature PL in mono- and bilayer MoTe$_2$ likely originates from similar mechanisms. However, the reduced PL quantum yield of bilayer MoTe$_2$ suggests that the indirect optical band gap is slightly smaller than the direct optical band gap such that phonon-assisted emission across the indirect optical band gap may contribute to the broadening of the PL spectrum in bilayer MoTe$_2$. Overall, we conclude that emission from the direct optical band gap dominates the room temperature PL response of bilayer MoTe$_2$.

Third, the PL spectra of $N\geq 3$-layer MoTe$_2$ differ markedly from the mono- and bilayer cases. We observe (i) a broad and prominent PL$^{-}$ feature (with a FHWM of approximately 100~meV), which, as $N$  increases, progressively dominates the narrower  PL$^+$ feature (with a FWHM in the range 60-70~meV), and (ii), as $N$ increases, PL$^-_\tr{max}$ downshifts significantly, while PL$^+_\tr{max}$ remains almost constant and very close to the energy of the A exciton absorption line~\cite{Ruppert2014}. In the bulk limit, the PL$^-$ peak is centered at 0.94~eV and is followed by a much fainter feature near 1.03~eV~\footnote{Following Refs.~\cite{Cappelluti2013,Brume2015} the difference between the values of the integrated PL intensities and of PL$^-_\tr{max}$ recorded in bulk and few-layer flakes ($N=6,7$) (see Fig.~\ref{Fig3} and Fig.~\ref{Fig4}) may arise from the fact that the bulk conduction band minimum occurs at a point in momentum space that lies halfway between the $\bm K$ and the $\bm \Gamma$ points, while the conduction band minimum is reached at the $\bm \Gamma$ point in the few-layer limit.}. Thus, the PL$^+$ and PL$^-$ peaks can tentatively be assigned to competing emission pathways, associated with hot luminescence from the A exciton and with phonon-assisted emission from the indirect excitons, respectively. Note that the PL$^-$ peak is broader than the PL$^+$ peak, presumably due to the phonons involved in the indirect emission process. Finally, our conclusions are further confirmed by the fact that the bulk values of PL$_\tr{max}^+$ and PL$_\tr{max}^-$ are in fair agreement with previous measurements of the bulk direct and indirect optical band gaps obtained from optical transmission spectroscopy~\cite{Lezama2014}.

\section{Exciton-exciton annihilation in monolayer \element{Mo}\element{Te}$_2$}

Having introduced monolayer MoTe$_2$ as a direct optical band gap semiconductor with bright near-infrared emission, we now focus on the influence of the exciton formation rate  $W_\tr{abs}$ on its PL quantum yield and PL spectral lineshape under cw laser excitation. $W_\tr{abs}$  is simply deduced from the \textit{effective} absorptance of monolayer MoTe$_2$ in the air/MoTe$_2$/SiO$_2$/Si layered structure, by taking into account the size of our tightly focused laser spot, the absorptance of bare MoTe$_2$~\cite{Ruppert2014} and optical interference effects (see Supplemental Material~\cite{SMnote}). For a laser photon energy of $2.33~\tr{eV}$, we calculated an absorptance of $\approx16.5~\%$ for monolayer MoTe$_2$ in our sample geometry. Assuming that one absorbed photon gives rise to one exciton, the exciton formation rates investigated here range from $W_\tr{abs}\approx 1.0\times 10^{19}\;\rm cm^{-2}~s^{-1}$ up to $3.6\times10^{22}\;\rm cm^{-2}~s^{-1}$.

\begin{figure}[!tb]
\begin{center}
\includegraphics[width=\linewidth]{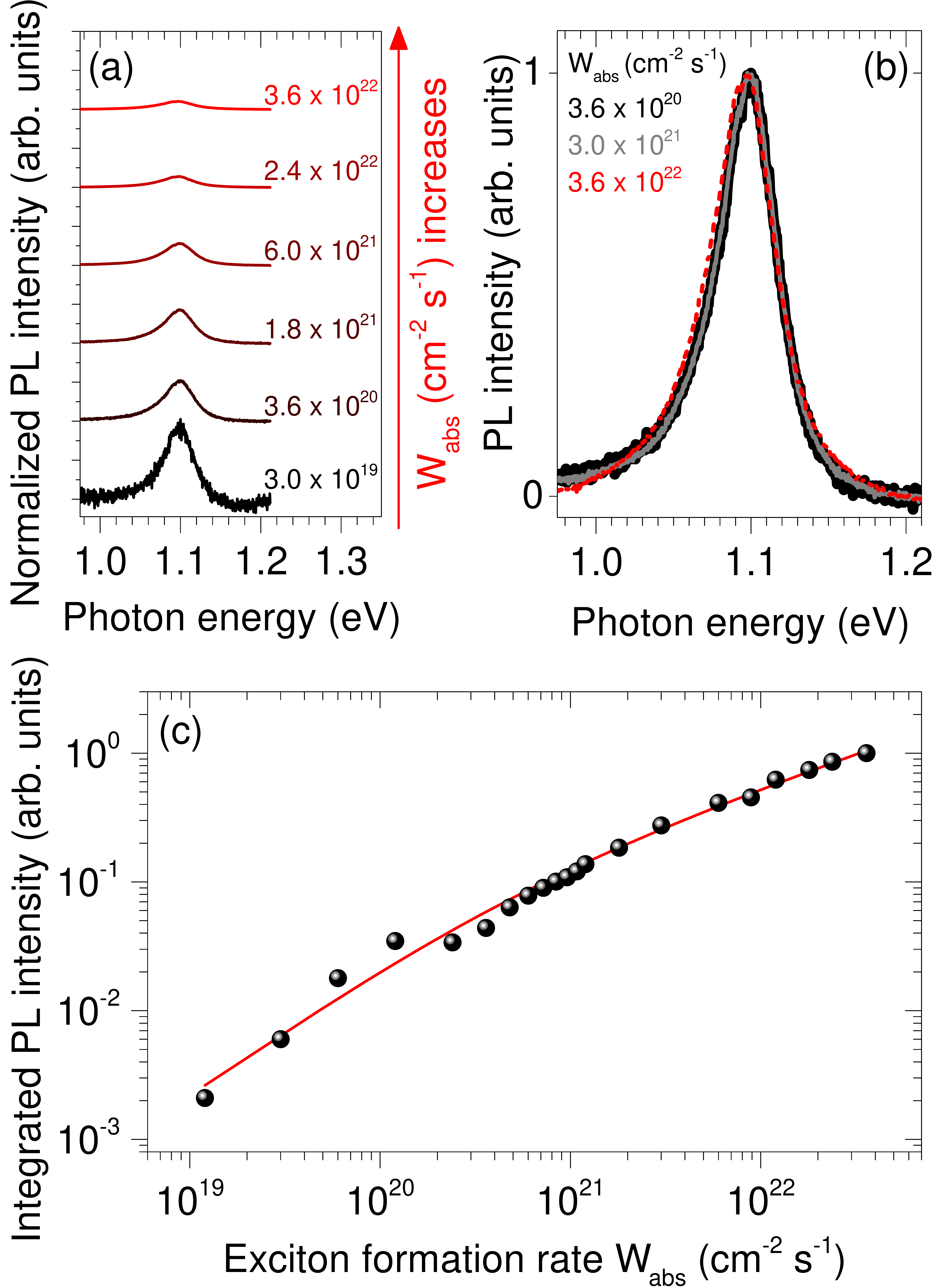}
\caption{(a) Photoluminescence spectra of a monolayer MoTe$_2$ sample at different exciton formation rates $W_\tr{abs}$. The spectra are normalized using the product of $W_\tr{abs}$ by the integration time and vertically offset for clarity. (b) Photoluminescence spectra of monolayer MoTe$_2$ for three different exciton formation rates. The spectra have been normalized to unity. (c) Integrated photoluminescence intensity obtained from the raw spectra (symbols) as a function of $W_\tr{abs}$ in monolayer MoTe$_2$. The solid line corresponds to a fit based on Eq.~\eqref{eq_nx}. The error bars are smaller than the symbol size.}
\label{Fig5}
\end{center}
\end{figure}



Figure~\ref{Fig5}(a,b) shows PL spectra recorded on the same monolayer for increasing values of $W_\tr{abs}$. The spectra have been normalized by the  incoming laser intensity (\textit{i.e.}, by $W_\tr{abs}$) and by the integration time. We clearly observe a non-linear decrease of the \textit{normalized} PL intensity that suggests, as shown in Fig.~\ref{Fig5}(c), that the \textit{raw} integrated PL intensity levels off with increasing $W_\tr{abs}$. We have checked that this non-linear behavior was not due to irreversible photo-induced damage of the sample~\cite{Chen2015b} and we have observed a very similar sub-linear rise of the PL intensity on another MoTe$_2$ monolayer (see Supplemental Material~\cite{SMnote}). As illustrated in Fig.~\ref{Fig5}(b), we notice that the linewidth of the PL spectra is independent of $W_\tr{abs}$ and that the PL spectra downshift very slightly (by only 3 meV) when $W_\tr{abs}$ reaches $3.6\times 10^{22}\;\rm cm^{-2} s^{-1}$ (\textit{i.e.}, a laser intensity of $81 ~\rm kW/cm^{2}$). We may thus conclude that biexciton emission~\cite{You2015} and photothermally-induced modifications of the PL spectra can be neglected for the range of exciton densities explored here. 


Sub-linear rises of the integrated PL intensity, as observed in Fig.~\ref{Fig5}(c), have recently been reported in other MX$_2$ monolayers (such as WSe$_2$~\cite{Mouri2014,Zhu2015,Yu2016}, WS$_2$~\cite{Zhu2015,Yuan2015,Yu2016}, or MoS$_2$~\cite{Yu2016}) and assigned to exciton-exciton annihilation (EEA). EEA has been further evidenced in these materials (and additionally in MoSe$_2$~\cite{Kumar2014}) by means of transient absorption spectroscopy~\cite{Kumar2014,Sun2014,Yu2016} or time-resolved PL measurements~\cite{Mouri2014,Yuan2015}, through the observation of accelerated exciton decays at high exciton densities. In order to further demonstrate our observation of EEA in monolayer MoTe$_2$, we make use of a simple rate equation model~\cite{Yu2016}. The integrated PL intensity is proportional to the steady state exciton density $\left\langle n_\tr{x}\right\rangle$. Assuming, that the time dependence of the exciton density $n_\tr{x}$ is essentially governed by the interplay between exciton formation (at a rate per unit area $W_\tr{abs}^{~}$), linear recombination (at a rate $\Gamma_\tr{x}$) and exciton-exciton annihilation (EEA) (at a rate $\gamma_\tr{eea}$), one obtains


\begin{equation}
\frac{\mathrm{d}n_\tr{x}}{\mathrm{d}t}= W_\tr{abs} - \Gamma_\tr{x} n_\tr{x} - \gamma_\tr{eea} n_\tr{x}^2.
\label{eq_rate}
\end{equation} 
The EEA term in this equation scales quadratically with $n_\tr{{x}}$ since the annihilation process involves Coulomb interaction between two excitons. The steady state exciton density is

\begin{equation}
\left\langle n_\tr{x}\right\rangle = \frac{\Gamma_\tr{x}}{2\gamma_\tr{eea}}\left(\sqrt{1+\frac{4\gamma_\tr{eea}}{\Gamma^2_\tr{x}}W_\tr{abs}}-1\right).
\label{eq_nx}
\end{equation}

The experimental data in Fig.~\ref{Fig5}(c) is very well fit by Eq.~\eqref{eq_nx}. From the fit, we extract $\gamma_\tr{eea}/\Gamma^2_\tr{x} \approx 1.4\times10^{-21}~\tr{cm}^2~\tr{s}$. Assuming a reasonable value of $\gamma_\tr{eea} \sim 0.1~\tr{cm}^2~\tr{s}^{-1}$, similar to previous estimates in substrate-supported MX$_2$ monolayers~\cite{Kumar2014,Sun2014,Yu2016,Mouri2014,Yuan2015}, one obtains a linear exciton recombination rate of $\Gamma_\tr{x}\sim 8.5\times10^9~\tr{s}^{-1}$, that is a room temperature exciton lifetime of $\sim 120~\tr{ps}$. Although additional near-infrared time-resolved measurements or transient absorption studies on monolayer MoTe$_2$ are needed to separately determine the exact values of $\gamma_\tr{eea}$ and $\Gamma_\tr{x}$, our simple analysis provides values that are in-line with recent room-temperature measurements on other MX$_2$~\cite{Yu2016,Robert2016}. Finally, let us also note that monolayer  MoTe$_2$ and related systems exhibit EEA rates that give rise to average exciton decay times similar to those reported in carbon nanotubes~\cite{Ma2005,Wang2004} in the non-linear regime. In addition, EEA  in MX$_2$ is much more efficient than related processes (\textit{i.e.}, Auger recombination) in conventional quantum wells~\cite{Sun2014,Haug1992,Taylor1996}. Highly efficient EEA between tighly bound excitons~\cite{Molina2013,Qiu2013,Ramasubramaniam2012,Chernikov2014,He2014,Ye2014,Wang2015,Zhu2015,Klots2014,Ugeda2014} in monolayer MX$_2$ reflects the strongly enhanced Coulomb interactions and reduced dielectric screening in these atomically thin two-dimensional materials.

\section{Conclusion and outlook}
We have performed a detailed analysis of the room temperature photoluminescence of $N$-layer MoTe$_2$. Monolayer MoTe$_2$ displays a direct optical band gap, with sharp emission at $1.10\rm~eV$. The crossover from a dominant direct excitonic emission (as observed in monolayers) to a dominant phonon-assisted indirect emission (in the bulk limit) occurs more smoothly than in other $2Hc$ transition metal dichalcogenides, such as MoS$_2$, MoSe$_2$, WS$_2$ and WSe$_2$. As a result, the difference between the bulk indirect optical band gap and the monolayer direct optical band gap is found to be only about $160~\tr{meV}$. Our observation of close-lying direct and indirect emission lines invites further calculations of exciton-phonon coupling in MoTe$_2$ and related systems, in order to correlate the values of the one-particle indirect band gap to the energy of the emission lines arising from indirect exciton recombination. Interestingly, in bilayer MoTe$_2$, the competition between direct and indirect emission may be efficiently manipulated by external electric fields~\cite{Ramasubramaniam2011,Zibouche2014,Brume2015}, in particular using dual-gated field effect transistors. In addition, we have unveiled a sub-linear scaling of the photoluminescence intensity of monolayer MoTe$_2$ with increasing exciton formation rate, which can be rationalized using a simple model based on exciton-exciton annihilation. This model also allowed us to obtain an order of magnitude estimate for the exciton lifetime in the linear regime that needs to be quantitatively confirmed by time-resolved photoluminescence measurements in the near-infrared range.





\begin{acknowledgments}

We are grateful to L. Wirtz and A. Molina-S\'anchez for stimulating discussions. We thank P. Bernhard at Roper Scientific for the loan of an InGaAs detector. We acknowledge financial support from the Agence Nationale de la Recherche (under grants QuanDoGra 12 JS10-001-01 and H2DH ANR-15-CE24-0016), from the LabEx NIE (Under grant WHO), from the CNRS and from Universit\'e de Strasbourg. 

\end{acknowledgments}




%


\onecolumngrid
\newpage
\begin{center}
{\LARGE\textbf{Supplemental Material}}
\end{center}

\setcounter{equation}{0}
\setcounter{figure}{0}
\setcounter{section}{0}
\renewcommand{\theequation}{S\arabic{equation}}
\renewcommand{\thefigure}{S\arabic{figure}}
\renewcommand{\thesection}{SI\,\arabic{section}}
\renewcommand{\thesubsection}{SI\,\arabic{section}\alph{subsection}}
\linespread{1.4}

\bigskip

\section{High-frequency Raman spectra}

\begin{figure}[!tbh]
\begin{center}
\includegraphics[width=0.75\linewidth]{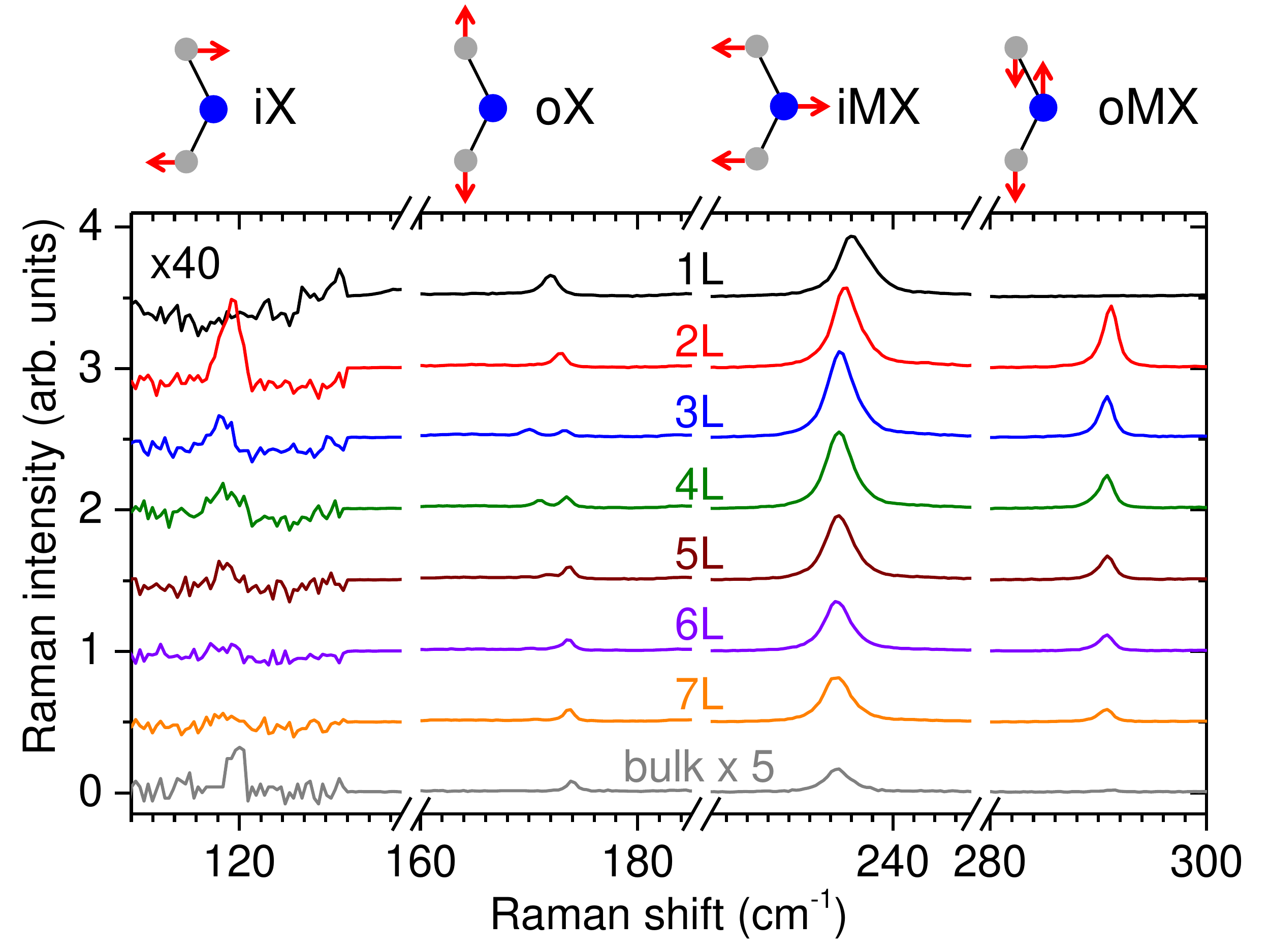}
\caption{High-frequency Raman spectra of $N=1$ to $N=7$ layers MoTe$_2$ and of bulk MoTe$_2$. The spectra are vertically offset for clarity. The four one-phonon features are labeled and the atomic displacements for the monolayer are indicated with Mo atoms in blue and Te atoms in grey.}
\label{FigSM1}
\end{center}
\end{figure}

Figure~\ref{FigSM1} shows the high-frequency range ($100 - 300~\tr{cm}^{-1}$) of the Raman spectra of Fig. 1(c) in the manuscript. We observe the four expected one-phonon features assigned to intralayer displacements in $2Hc$ transition metal dichalcogenides~\cite{Zhang2015,Froehlicher2015}: (i) the in-plane, out-of-phase vibration of the Te planes, with E$_{1g}$ symmetry in bulk (iX mode at $120~\tr{cm}^{-1}$), (ii) the out-of-plane, out-of-phase vibration of the Te planes, with A$_{1g}$ symmetry in bulk (oX mode at $170~\tr{cm}^{-1}$), (iii) the in-plane vibration of the Mo and Te planes against each other, with E$_{2g}$ symmetry in bulk (iMX mode at $235~\tr{cm}^{-1}$), and (iv) the out-of-plane vibration of the Mo and Te planes against each other, with B$_{2g}$ symmetry in bulk (oMX mode at $290~\tr{cm}^{-1}$). Note that we discern the Davydov splitting of the oX mode as recently reported in Refs.~\cite{Froehlicher2015,Grzeszczyk2016,Song2016}, confirming the number of layers $N$ deduced from the low-frequency part of the spectra. 

\clearpage

\section{Laser spot area}

\begin{figure}[!tbh]
\begin{center}
\includegraphics[width=0.8\linewidth]{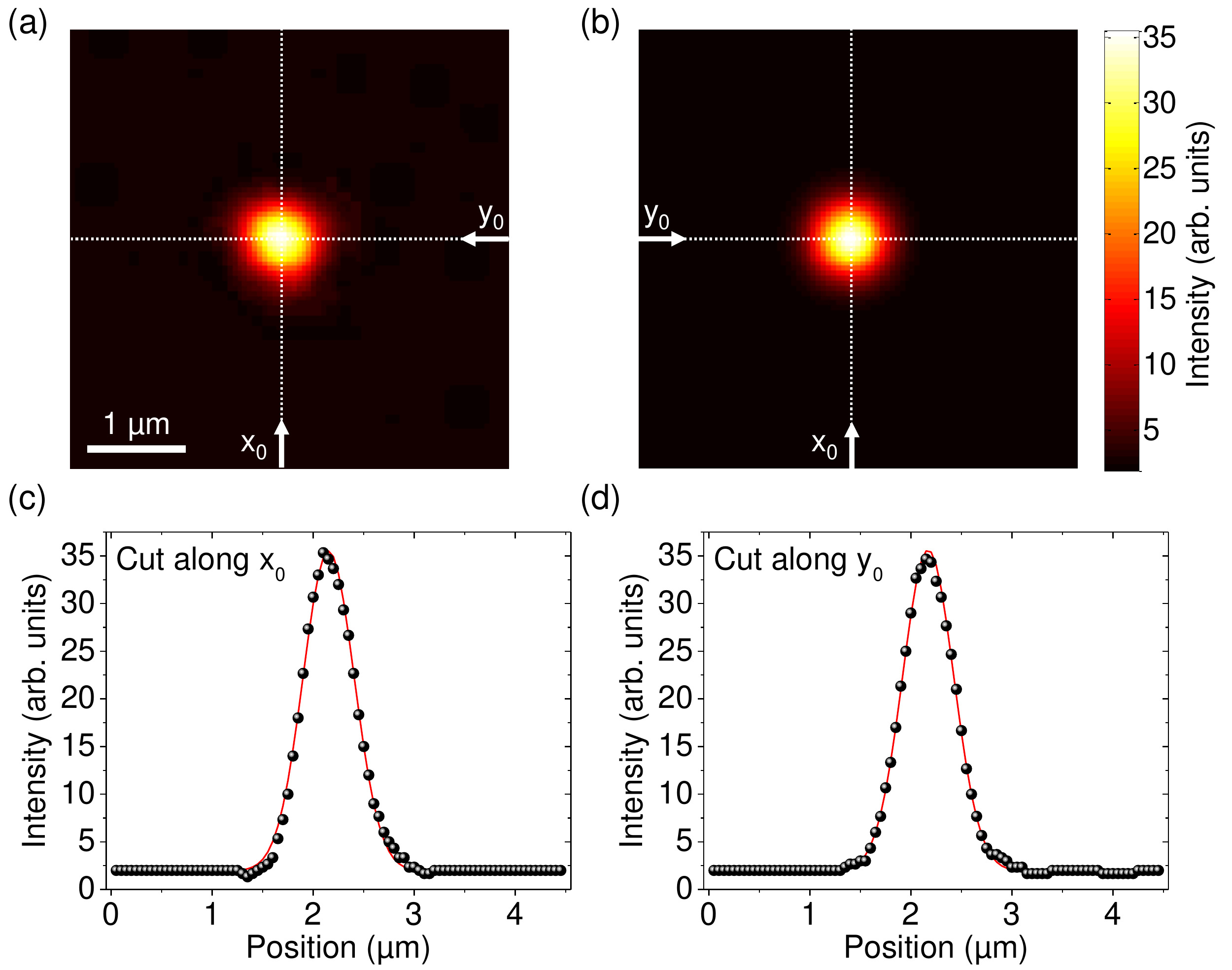}
\caption{(a) Optical image of the focused laser spot on the surface of a substrate. (b) Two-dimensional fit of the optical image. Cuts along (c) $x_0$ and (d) $y_0$. The solid lines are the fit to the experimental data (symbols).}
\label{FigSM2}
\end{center}
\end{figure}

In order to measure the area of our laser spot, we recorded an optical image of the tightly focused laser spot on the surface of a substrate (see Fig.~\ref{FigSM2}(a)). We have then fitted this image with a two-dimensional Gaussian function
\begin{equation}
f(x,y) = A \exp\left(- \frac{(x-x_o)^2+(y-y_o)^2}{2\sigma^2}\right),
\end{equation}
where $A$ is the amplitude of the Gaussian, $(x_0, y_0)$ are the coordinates of the center and $\sigma$ is the standard deviation (we assumed that the standard deviation is the same for the two dimensions). On Fig.~\ref{FigSM2}(b)-(d), we observe that the data are well fitted by this function. Knowing that the surface area is given by $2\pi\sigma^2$, we deduced a laser spot area of $4\times10^{-9}~\tr{cm}^2$ that has been used to estimate the exciton formation rate per unit area $W_\tr{abs}$ in the main manuscript.

\clearpage

\section{Interference effects}

It is well-known that interference effects strongly affect optical absorption, as well as the Raman~\cite{Yoon2009,Li2012b} and PL~\cite{Buscema2014} signal of layered materials. Indeed, multiple reflections at different interfaces (air/MoTe$_2$, MoTe$_2$/SiO$_2$ and SiO$_2$/Si) can enhance the absorption of the incoming light beam (see Fig.~\ref{FigSM3} and Fig.~\ref{FigSM4}(a)) as well as the PL (or Raman) intensity (see Fig.~\ref{FigSM4}(b)) by a factor F$_\tr{ab}$ and F$_\tr{pl}$, respectively.

\subsection{Absorptance of the monolayer MoTe$_2$}

\begin{figure*}[!tbh]
\begin{center}
\includegraphics[width=0.5\linewidth]{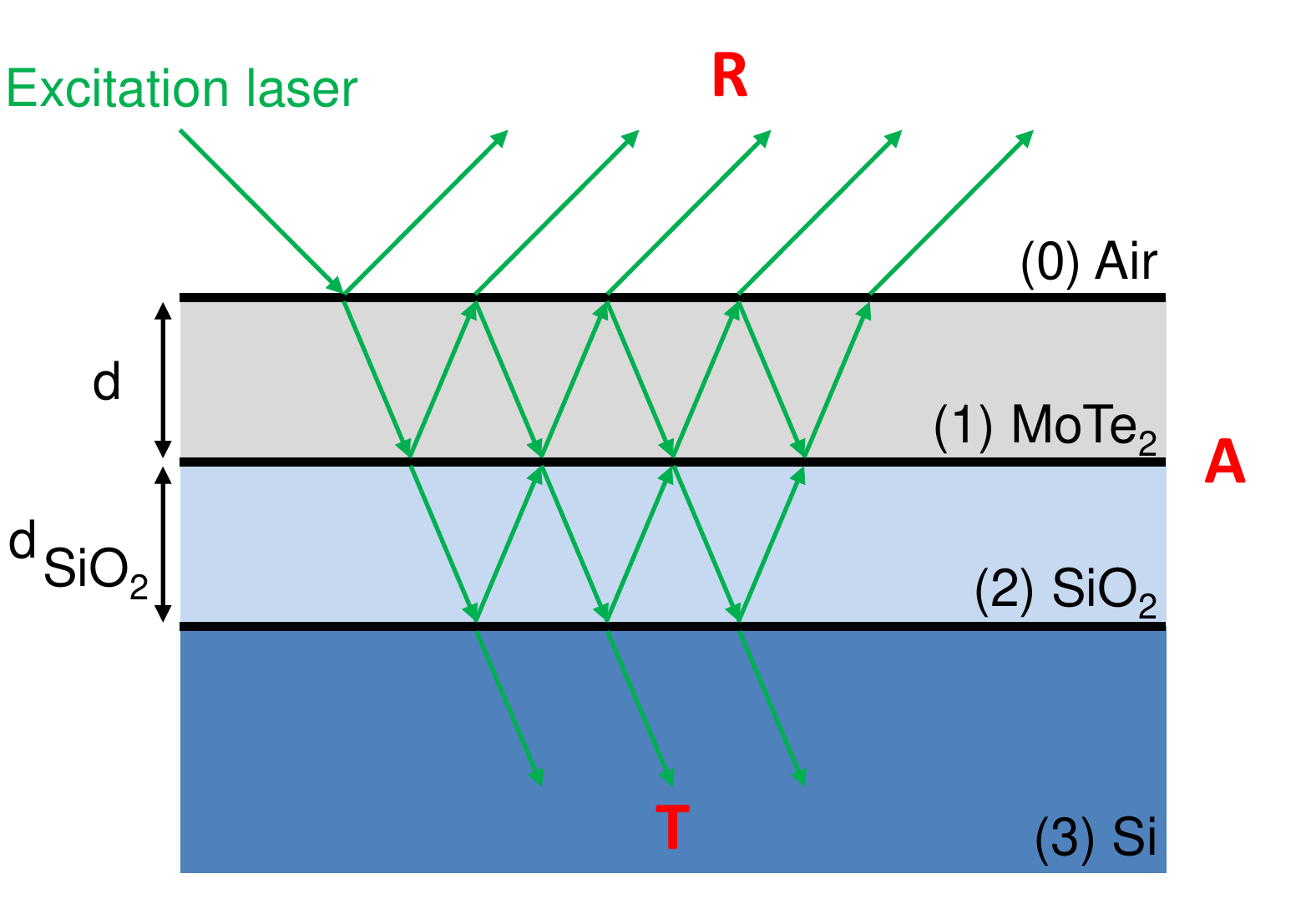}
\caption{Schematic diagrams of the optical paths in our geometry for a laser photon energy of $2.33~\tr{eV}$. $R$ is the reflectance, $T$ the transmittance and $A$ the absorptance.}
\label{FigSM3}
\end{center}
\end{figure*}

We first consider the absorptance of $N$-layer MoTe$_2$. Due to interference effects in the air/MoTe$_2$/SiO$_2$/Si structure, the absorptance $A$ (absorbed fraction of incident light) of $N$-layer MoTe$_2$ is different as compared to the freestanding case. As it is drawn in Fig.~\ref{FigSM3}, a fraction $R$ (reflectance) of the incident excitation light is reflected, a fraction $T$ (transmittance) is transmitted into the Si substrate, which is supposed to be semi-infinite (in practice the substrate only needs to be ticker than a few absorption lengths) and a fraction A is absorbed. Energy conservation imposes \cite{Hecht2001} 
\begin{equation}
A + R + T = 1.
\end{equation}

Since SiO$_2$ is supposed to be transparent, the absorptance of $N$-layer MoTe$_2$ in this structure is directly given by $A=1-R-T$. $R$ and $T$ can be obtained analytically using an interference calculation~\cite{Yoon2009,Li2012b}. Without changing drastically the results, we can assume that the light impinges on the sample at normal incidence. In this condition, the Fresnel coefficients are $t_{ij}=2n_i/(n_i+n_j)$ and $r_ {ij}=(n_i-n_j)/(n_i+n_j)$. $n_0=1$ is the refractive index of air and $n_1$, $n_2$ and $n_3$ are the complex refractive index for MoTe$_2$, SiO$_2$ and Si, respectively. The phase factors are $\beta_1=2\pi n_1 d/\lambda$ and $\beta_2=2\pi n_2 d_{\tr{SiO}_2}/\lambda$ where $\lambda$ is the wavelength of the light in vacuum and $d$ ($d_{\tr{SiO}_2}$) is the thickness of MoTe$_2$ (SiO$_2$). Thus, the normal incidence reflectance $R$ and transmittance $T$ are given by \cite{Hecht2001}
\begin{subequations}
\begin{align}
R&=\abs{\frac{r_{01}[1+r_{12}r_{23}e^{2i\beta_2}]+[r_{12}+r_{23}e^{2i\beta_2}]e^{2i\beta_1}}{1+r_{12}r_{23}e^{2i\beta_2}+[r_{12}+r_{23}e^{2i\beta_2}]r_{01}e^{2i\beta_1}}}^2,\\
T&=\abs{\frac{t_{01}t_{12}t_{23}t_{32}t_{21}t_{10}e^{2i(\beta_1+\beta_2)}}{(1+r_{12}r_{23}e^{2i\beta_2}+[r_{12}+r_{23}e^{2i\beta_2}]r_{01}e^{2i\beta_1})^2}}.
\end{align}
\end{subequations}
At a photon energy of $2.33~\tr{eV}$, the refractive index of MoTe$_2$ is $n_{2.33~\tr{eV}}=4.07+1.63i$~\cite{Beal1979}, of SiO$_2$ $n_{\tr{SiO}_2}=1.4607$~\cite{Malitson1965} and of Si $n_\tr{Si}=4.14+0.045i$~\cite{Palik1998}. For a monolayer MoTe$_2$ of thickness $c/2=0.6984~\tr{nm}$~\cite{Boker2001} and a SiO$_2$ layer of $d_{\tr{SiO}_2}=90~\tr{nm}$, we calculated an absorptance $A\approx16.5~\%$, which is in line with the absorptance of other TMDs measured on SiO$_2$/Si~\cite{Tonndorf2013}.

\clearpage

\subsection{Normalization process}

 In order to quantitatively compare the PL spectra recorded on $N$-layer MoTe$_2$, one has to take into account (i) the response of the setup (grating and camera) and (ii) the optical interference and absorption effects. The response of the camera is supposed to be flat (it varies by less than $5\%$) in the spectral range studied here. The response of the grating is extracted from the data of the manufacturer (Richardson Gratings 53-*-500R).

\begin{figure*}[!ht]
\begin{center}
\includegraphics[width=0.65\linewidth]{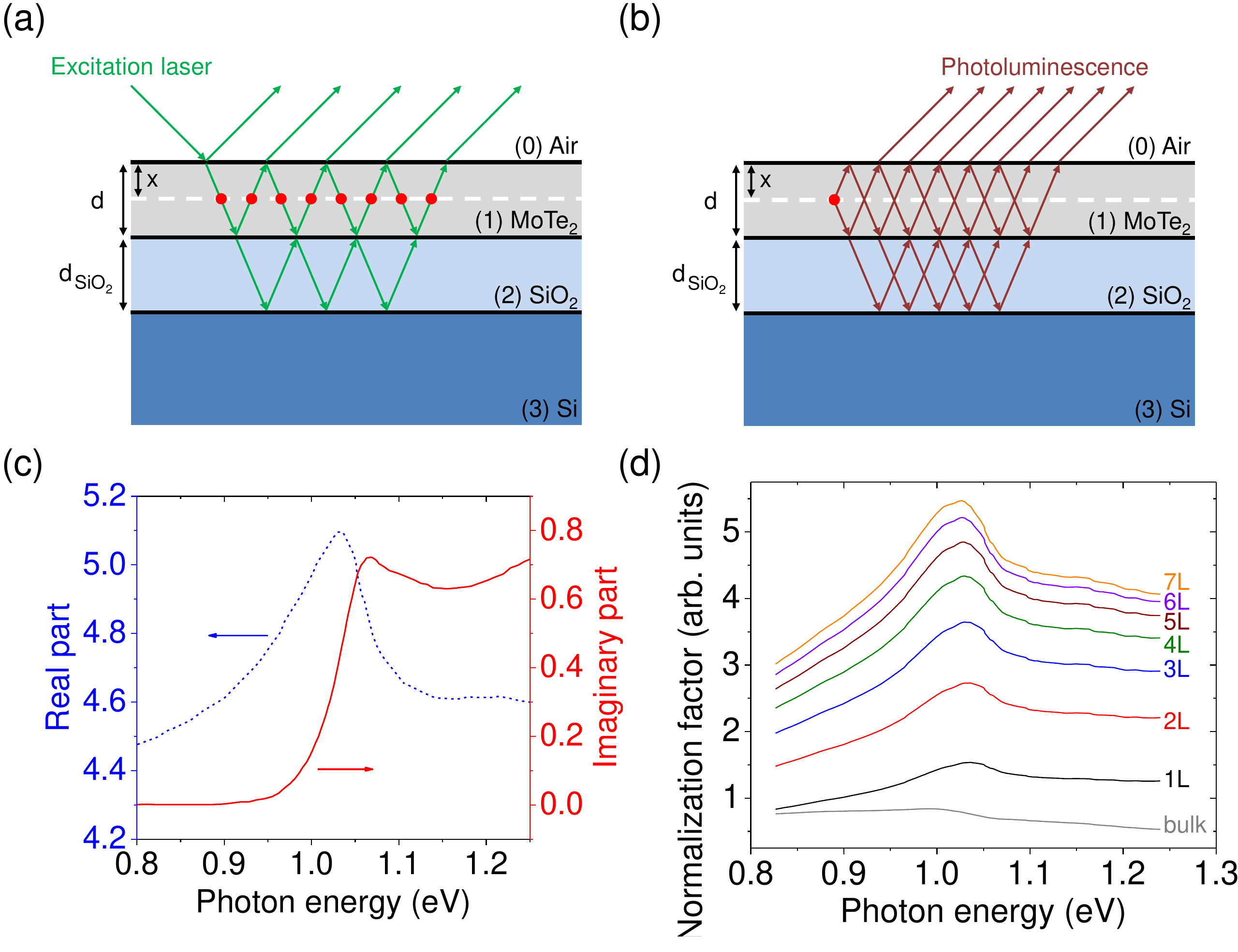}
\caption{Schematic diagram of the optical paths in our geometry for (a) the excitation laser (at a photon energy of $2.33~\tr{eV}$) and (b) the emitted light. $d$ is proportional to $N$.(c) Refractive index of MoTe$_2$, extracted from the measurements in Ref.~\cite{Lezama2014}, as a function of the emission energy. (d) Normalization factor as function of the emission energy.}
\label{FigSM4}
\end{center}
\end{figure*}

Following the results reported in Refs.~\cite{Yoon2009,Li2012b}, we calculated the enhancement factor for the PL due to the multiple interference

\begin{equation}
F=\int_{0}^{d} \abs{F_\tr{ab}(x)F_\tr{pl}(x)}^2 \, \mathrm{d}x,
\end{equation}
with 
\begin{subequations}
\begin{align}
       F_\tr{ab}&=t_{01}\frac{[1+r_{12}r_{23}e^{2i\beta_2}]e^{i\beta_x}+[r_{12}+r_{23}e^{2i\beta_2}]e^{i(2\beta_1-\beta_x)}}{1+r_{12}r_{23}e^{2i\beta_2}+[r_{12}+r_{23}e^{2i\beta_2}]r_{01}e^{2i\beta_1}},\\
       F_\tr{pl}&=t_{10}\frac{[1+r_{12}r_{23}e^{2i\beta_2}]e^{i\beta_x}+[r_{12}+r_{23}e^{2i\beta_2}]e^{i(2\beta_1-\beta_x)}}{1+r_{12}r_{23}e^{2i\beta_2}+[r_{12}+r_{23}e^{2i\beta_2}]r_{01}e^{2i\beta_1}},
\end{align}
\end{subequations}
where we used exactly the same notations as previously. $\beta_x=2\pi n_1 x/\lambda$ with $x$ being the depth of the point where the interactions occur (see Fig.~\ref{FigSM4}(a)-(b)). Note that the expressions for F$_\tr{ab}$ and F$_\tr{pl}$ are similar, but the wavelengths in the phase factors are different.

For the excitation laser, we used the same refractive index as for the absorptance calculations. For the PL, we used the values extracted from the dielectric function measured in Ref.~\cite{Lezama2014} (see Fig.~\ref{FigSM4}(c)) for MoTe$_2$, and the tabulated values for SiO$_2$ and Si from Refs.~\cite{Malitson1965} and \cite{Palik1998}, respectively.
We also used $d=N\times c/2$, where $c/2=0.6984~\tr{nm}$~\cite{Boker2001} is the thickness of one layer. 

By multiplying the enhancement factors and the grating response, we obtained the normalized factors plotted in Fig.~\ref{FigSM4}(d). Note that for the bulk, we supposed that MoTe$_2$ is semi-infinite, \textit{i.e.}, there is only one interface, air/MoTe$_2$. Finally, we divided the PL spectra by the corresponding normalization factor.

\section{PL spectra at two different photon energies}

\begin{figure}[!tbh]
\begin{center}
\includegraphics[width=0.5\linewidth]{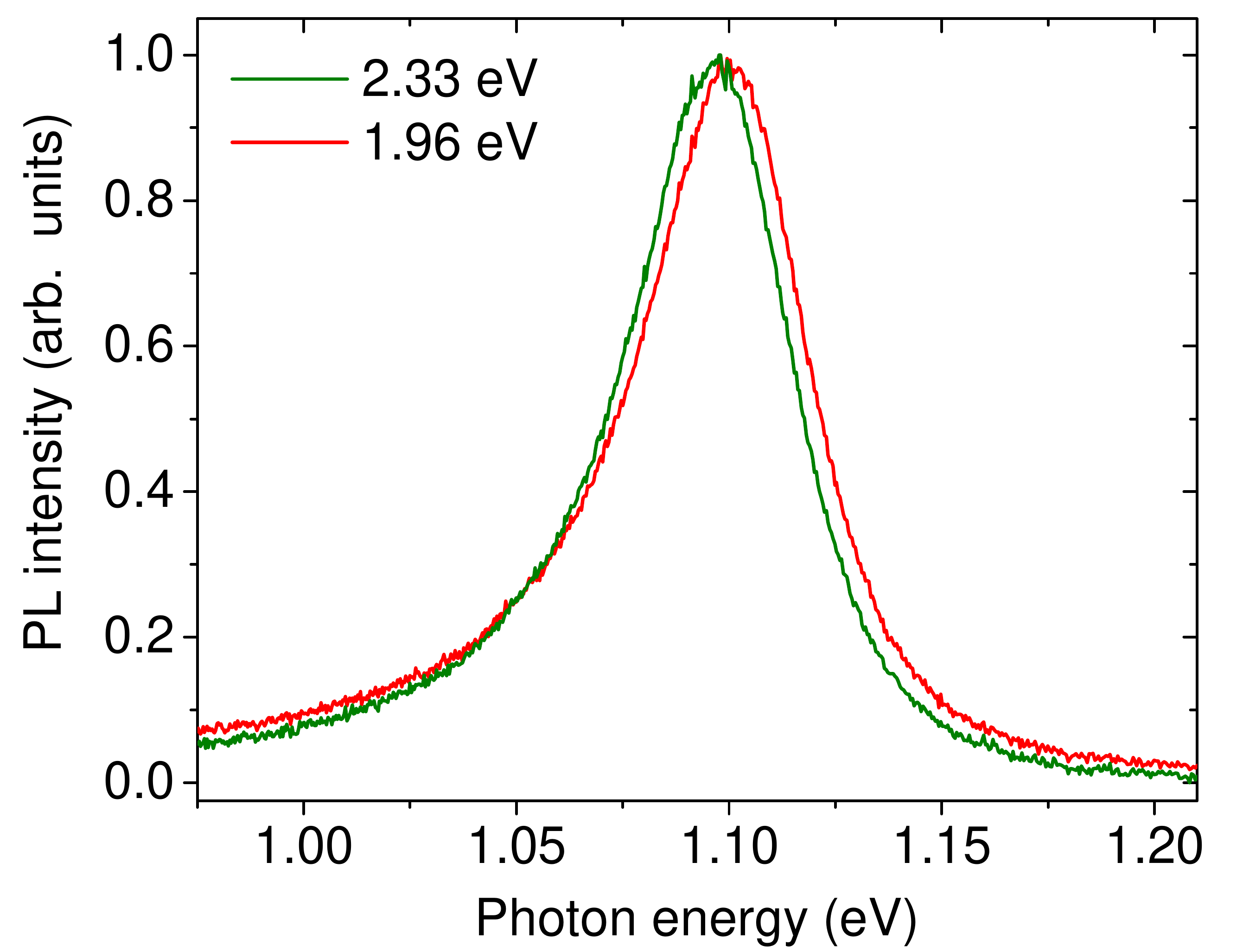}
\caption{Photoluminescence spectra of a monolayer MoTe$_2$ recorded under the same condition at laser photon energies of $2.33~\tr{eV}$ and $1.96~\tr{eV}$. When exciting at $1.96~\tr{eV}$, we observe that the emission energy blueshifts by $\approx 6~\tr{meV}$ compared to a reference spectrum recorded using a laser excitation at $2.33~\tr{eV}$. However, we do not observe significant modifications of the PL lineshape.}
\label{FigSM5}
\end{center}
\end{figure}



\section{Photoluminescence of the Si substrate}

\begin{figure}[!ht]
\begin{center}
\includegraphics[width=0.5\linewidth]{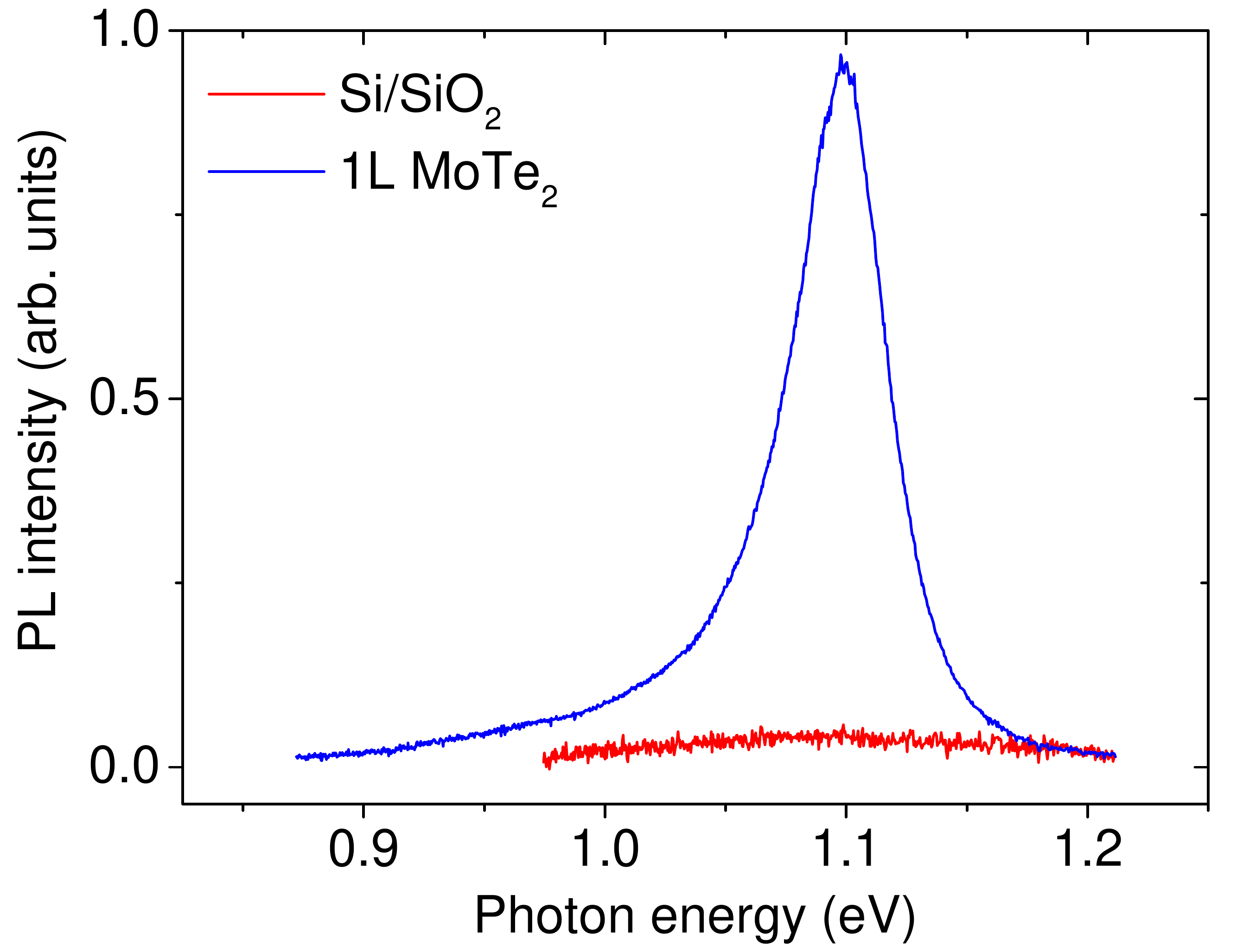}
\caption{Photoluminescence spectra of a monolayer MoTe$_2$ and of the bare Si/SiO$_2$ substrate recorded at $2.33~\tr{eV}$ under a laser intensity of $6.5\rm~kW~cm^{-2}$ (corresponding to an exciton formation rate of $W_\tr{abs}\approx2.9\times10^{21}~\tr{cm}^{-2}~\tr{s}^{-1}$ in monolayer MoTe$_2$). The emission from the substrate is negligible in our study.}
\label{FigSM6}
\end{center}
\end{figure}

\clearpage

\section{Exciton-exciton annihilation on another sample}

\begin{figure}[!tbh]
\begin{center}
\includegraphics[width=0.6\linewidth]{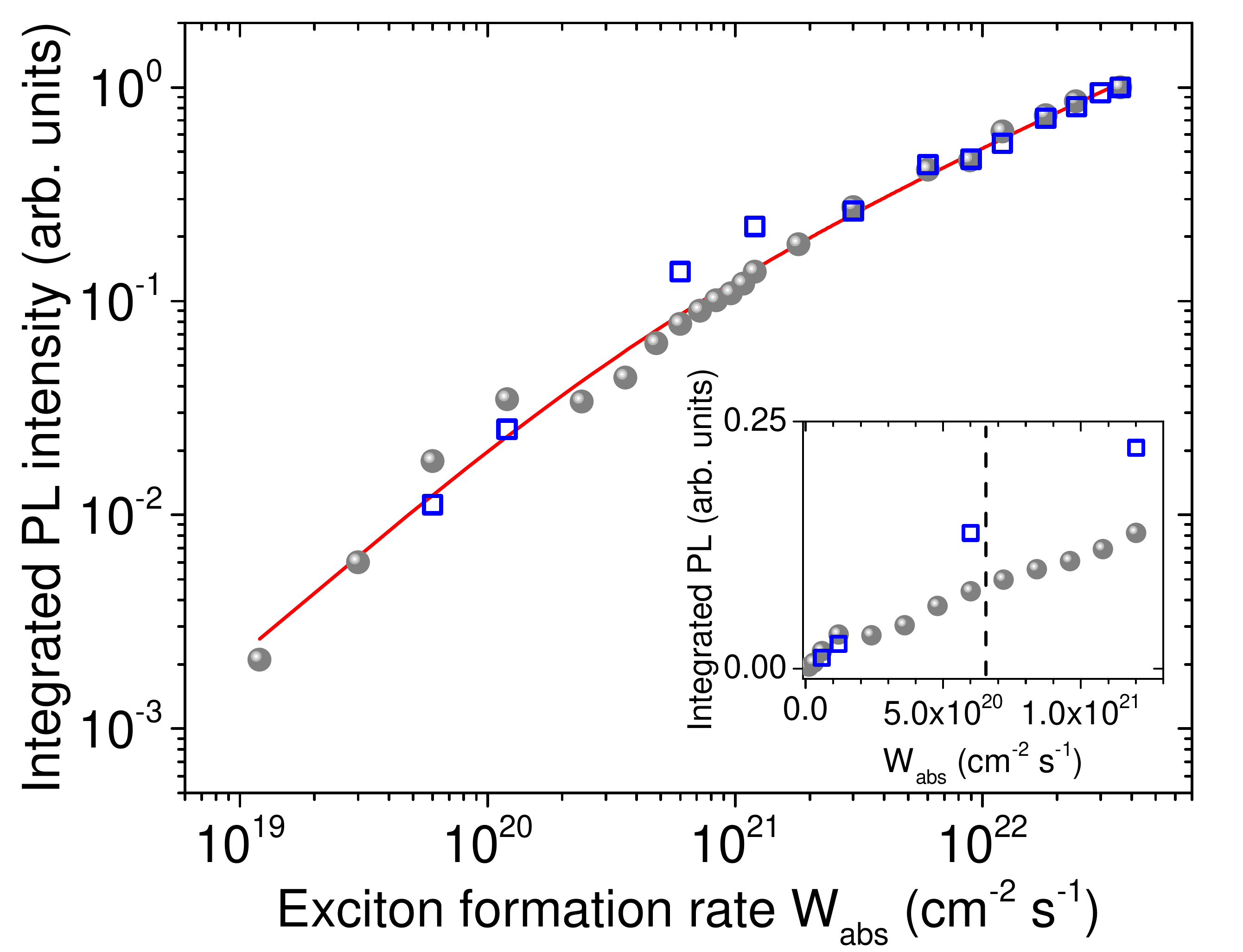}
\caption{Integrated photoluminescence intensity as a function of the exciton formation rate $W_\tr{abs}$ for two different monolayer samples of MoTe$_2$. The gray-filled circles are the same data as Fig.~5(c) in the manuscript  (sample 1). The solid red line is the fit to this data using Eq. (2), as in the manuscript. The raw PL intensity from the second sample (sample 2) has been multiplied by a factor of 2.4 (open blue squares) to show a clearer comparison with the data from sample 1. The normalized PL intensity from sample 2 scales very similarly as the PL intensity from sample 1, suggesting similar exiton-exciton annihilation rates and linear exciton decay rates for both samples. The inset shows the same data at low exciton formation rates $W_{\rm abs}$, on a linear scale. The vertical dashed line indicates the excitation formation rate at which the measurements on $N$-layer MoTe$_2$ shown in Fig. 2 and 3 of main manuscript have been performed. The error bars are smaller than the symbol size.}
\label{FigSM7}
\end{center}
\end{figure}

\end{document}